\documentclass{elsart}
\usepackage{epsfig}
 \usepackage{graphicx}
\usepackage{epstopdf}
\usepackage{placeins}

\usepackage{amssymb}
\journal{Physics Letters B}
\def\v#1{\mbox{\boldmath$#1$}}
\def\ket#1{|#1 \rangle}
\def\bra#1{\langle #1|}

\newcommand{\mc}{\multicolumn}
\newcommand{\lsim}{\mathrel{\mathop{\kern 0pt \rlap
  {\raise.2ex\hbox{$<$}}}
  \lower.9ex\hbox{\kern-.190em $\sim$}}}
\newcommand{\gsim}{\mathrel{\mathop{\kern 0pt \rlap
  {\raise.2ex\hbox{$>$}}}
  \lower.9ex\hbox{\kern-.190em $\sim$}}}

\begin{document}
\begin{frontmatter}

\title{Nonmesonic Weak Decay of $\Lambda$ Hypernuclei:
The Three--Nucleon Induced Mode}

\author{E.~ Bauer$^{1,2}$, G.~Garbarino$^3$ and C.~ A.~ Rodr\'{\i}guez Pe\~na$^2$}

\address{$^1$Facultad de Ciencias Astron\'omicas y Geof\'{\i}sicas, \\
UNLP, 1900-La Plata, Argentina}

\address{$^2$ IFLP, CONICET, 1900-La Plata, Argentina}

\address{$^3$IIS G. Peano, I-10147 Torino, Italy}

\date{\today}

\begin{abstract}
The nonmesonic weak decay of $\Lambda$ hypernuclei is studied within a microscopic diagrammatic
approach which is extended to include the three--nucleon induced mechanism.
We adopt a nuclear matter formalism which, through the local density approximation,
allows us to model finite hypernuclei, a one--meson--exchange weak transition potential
and a Bonn nucleon--nucleon strong potential. One--, two-- and three--nucleon induced weak
decay rates are predicted for $^{12}_\Lambda$C by including ground state correlations up to
second order in the nucleon--nucleon potential and the recoil of the residual nucleus.
Three--nucleon stimulated decays, $\Lambda NNN\to nNNN$ ($N=n$ or $p$), are considered
here for the first time. The obtained decay rates compare well with the latest KEK and
FINUDA data. The three--nucleon induced rate turns out to be dominated by $nnp$--
and $npp$--induced decays, it amounts to  $\sim$ 7\% of the total
nonmesonic rate and  it is $\sim 1/2$ of the neutron--induced decay rate.
The reduction effect of the nuclear recoil is particularly relevant for the
three--nucleon induced rates ($\sim$ 15\%), less important for the two--nucleon induced rates
($\sim$ 4\%) and negligible for the one--nucleon induced rates. Given the non--negligible size
of the three--nucleon induced contribution and consequently its importance in the precise
determination of the complete set of decay rates, new measurements and/or experimental
analysis are encouraged.
\end{abstract}


\end{frontmatter}

\newpage

{\bf \em Introduction} --
Since the first observation of $\Lambda$ hypernuclei in 1953 and the introduction of the strangeness quantum number in the same year, strange nuclei have been investigated with increasing theoretical and experimental efforts \cite{Rev-hypernuclei}. Hypernuclear physics is nowadays a mature field of research which in many aspects is located at the crossroads between particle and nuclear physics. It implies important connections with QCD \cite{Br14} ---consider the relevance of the production of hypernuclei and anti--hypernuclei in relativistic heavy--ion collisions and the possible extension of the usual techniques of lattice QCD, effective field theories and chiral perturbation theory to the baryon--baryon interactions in the strange sector--- as well as with astrophysical processes and observables \cite{Stars}, where it provides important inputs to study the thermal evolution, the stability, the macroscopic properties and the composition of compact astrophysical objects, including the so--called ``hyperon puzzle'' in neutron stars.

Production, structure and decays are the available tools to investigate those properties of hypernuclei which in turn allow us to have access to the elementary weak and strong hyperon interactions within the flavor $SU(3)$ sector \cite{Rev-hypernuclei}; the amount of information needed to this end would be practically impossible to obtain in scattering experiments.

A $\Lambda$--hypernucleus weakly decays to non--strange final states by mesonic ($\Lambda \to \pi^-p$ and $\Lambda \to \pi^0 n$) and nonmesonic modes (one--nucleon induced, $\Lambda N\to nN$, two--nucleon induced, $\Lambda NN\to nNN$, etc, $N$ denoting either a neutron or a proton). A satisfactory theory--experiment agreement has been reached for the mesonic weak decay and allowed us to test the proposed pion--nucleus optical potentials.
Most of the efforts have been devoted to the study of the nonmesonic weak decay, which from the beginning in the seventies posed important and subtle questions.

Most of the long--standing problems concerning the nonmesonic weak decay, which consisted in severe disagreements between theoretical and experimental predictions, have been solved in the last ten years or so \cite{Rev-hypernuclei,Rev-decay}: we mention the ``$\Gamma_n/\Gamma_p$ puzzle'' on the ratio between the decay rates for the neutron-- and the proton--induced processes, $\Lambda n \to nn$ and $\Lambda p \to np$, and the ``asymmetry puzzle'' in the decay of polarized hypernuclei. Different measurements of single and coincidence spectra of the nucleons emitted in hypernuclear nonmesonic decay, by the SKS Collaboration at KEK \cite{KEK-spectra,KEK09} and the FINUDA Collaboration at LNF \cite{FINUDA,FINUDA3,FINUDA4}, their theoretical analysis and the many calculations incorporating: 1) complete meson--exchange weak interaction potentials (with both one-- and two--meson exchange) \cite{Pa97,Ji01,It10,Kr10}, 2) the description of the $\Lambda N$ and $NN$ short--range correlations in terms of quark degrees of freedom \cite{Sa05}, 3) two--nucleon induced modes, $\Lambda NN \to nNN$ \cite{al91,Ra94,Ga00, Al00, BG09,BG10}, as well as 4) nucleon final state interactions \cite{GPR03,BGPR10,BG12}, have proven crucial to solve the ``$\Gamma_n/\Gamma_p$ puzzle'', in a continuous dialogue and exchange between theory and experiment.
The two--nucleon induced decay mechanism is rather well understood at present too, although the data on $\Gamma_2$ are still affected by large error bars.
Concerning the asymmetry parameter in the nonmesonic decay of polarized hypernuclei, we briefly recall that theoretical models including the exchange of uncorrelated and correlated pion pairs (in addition to one--meson exchanges) and nucleon final state interactions \cite{BGPR12} solved the ``asymmetry puzzle'' by reproducing the small asymmetry values measured at KEK \cite{Ma07}.

As mentioned, the present agreement between theory and experiment on the nonmesonic decay rates and asymmetries of $s$-- and $p$--shell hypernuclei is rather good, although the relatively large experimental error bars do not allow us to discriminate between the various proposed weak interaction schemes. One should remember that involved nuclear many--body processes such as multinucleon induced nonmesonic weak decays and nucleon final state interactions due to the nucleon--nucleon strong interaction complicate the extraction of the relevant elementary information. Discrepancies between theory and experiment indeed remain and mainly concern proton emission, i.e., proton kinetic energy spectra and neutron--proton angular and momentum correlation spectra \cite{Rev-decay,BG12}. A detailed description of the hyperon weak interactions underlying hypernuclear decay is thus impossible at present. From the theoretical viewpoint, new decay mechanisms should be explored.

An aspect to be investigated concerns the possible relevance of multinucleon induced decay mechanisms beyond the two--nucleon induced one.
The aim of the present work is to present the first evaluation of the rates for the three--nucleon induced nonmesonic weak decays, $\Lambda NNN \to nNNN$,
which proceed through the following isospin channels: $\Lambda nnn \to nnnn$, $\Lambda nnp \to nnnp$, $\Lambda npp \to nnpp$, and $\Lambda ppp \to nppp$. The effect of the nuclear recoil is also taken into account as a new feature of our approach, which is applied to $^{12}_\Lambda$C. Formally, this corresponds to implement translational invariance and exact momentum conservation in the nonmesonic decays.
The same microscopic approach showed that
ground state correlation contributions are crucial for a detailed calculation of the
rates, the asymmetry parameter and the nucleon emission spectra
 \cite{BG09,BG10,BGPR10,BGPR12}. Less pronounced effects were reported by including the $\Delta$--baryon resonance \cite{BG12}.

{\bf \em Formalism} --
Let us start the discussion of our formalism by establishing the notation adopted for the decay rates. The total nonmesonic decay rate reads:
\begin{equation}
\Gamma_{\rm NM}=\Gamma_1+\Gamma_2+\Gamma_3 \, ,
\end{equation}
where $\Gamma_1$, $\Gamma_2$ and $\Gamma_3$ denote the rates for the one--, two-- and three--nucleon stimulated decays, $\Lambda N\to nN$, $\Lambda NN\to nNN$ and $\Lambda NNN\to nNNN$, where $N=n$ or $p$.
In terms of the various isospin channels, the one--, two-- and three--nucleon induced nonmesonic rates are decomposed as:
\begin{eqnarray}
\Gamma_1&=& \Gamma_n+\Gamma_p \, , \\
\Gamma_2&=& \Gamma_{nn}+\Gamma_{np} +\Gamma_{pp}\, , \\
\Gamma_3&=& \Gamma_{nnn}+\Gamma_{nnp} +\Gamma_{npp} +\Gamma_{ppp}\, .
\end{eqnarray}
The sub--indices on the rhs expressions indicate the initial (multi)nucleon state stimulating the weak decay: $\Gamma_{nnp}\equiv \Gamma(\Lambda nnp\to nnnp)$, etc.

The analytical expressions for the nonmesonic decay widths are derived as follows.
To obtain the one--, two-- and three--nucleon induced rates for a $\Lambda$ with four--momentum $k=(k_0,\v{k})$ inside infinite nuclear matter with Fermi momentum $k_F$, we start by the partial widths:
\begin{equation}
\label{decw}
\Gamma_{1 \, (2, \, 3)}(k,k_{F}) = \sum_{f} \,
 |\bra{f} V^{\Lambda N\to nN} \ket{0}_{k_{F}}|^{2}
\delta (E_{f}-E_{0})~,
\end{equation}
where $\ket{0}_{k_{F}}$ and $\ket{f}$ are the initial
hypernuclear ground state (whose energy is $E_0$)
and the possible $2p1h$, $3p2h$ or $4p3h$ final states (with energy $E_f$), respectively.
The $2p1h$ ($3p2h$, $4p3h$) final states define the rate $\Gamma_{1}$ ($\Gamma_{2}$, $\Gamma_{3}$), while $V^{\Lambda N\to nN}$ is the two--body weak transition potential. This potential contains the exchange of the full
set of mesons of the pseudoscalar ($\pi$, $\eta$, $K$) and vector octets ($\rho$, $\omega$, $K^*$), with strong coupling constants and cut--off
parameters deduced from the Nijmegen soft--core interaction NSC97f \cite{St99} (for details on the weak transition potential we refer to \cite{Pa97}).

The rates for a finite hypernucleus are obtained by the local density approximation~\cite{Os85}, i.e., after averaging the above partial widths over the $\Lambda$
momentum distribution in the considered hypernucleus,
$|\widetilde{\psi}_{\Lambda}(\v{k})|^2$, and over the local Fermi momentum,
$k_{F}(r) = \{3 \pi^{2} \rho(r)/2\}^{1/3}$,
$\rho(r)$ being the density profile of the hypernuclear core.
One thus has:
\begin{equation}
\label{decwpar3}
\Gamma_{1 \, (2, \, 3)} = \int d \v{k} \, |\widetilde{\psi}_{\Lambda}(\v{k})|^2
\int d \v{r} \, |\psi_{\Lambda}(\v{r})|^2
\Gamma_{1 \, (2, \, 3)}(\v{k},k_{F}(r))~,
\end{equation}
where for $\psi_{\Lambda}(\v{r})$, the Fourier transform of
$\widetilde{\psi}_{\Lambda}(\v{k})$, we
adopt the $1s_{1/2}$ harmonic oscillator wave--function with
frequency $\hbar \omega$ ($=10.8$ MeV for $^{12}_\Lambda$C)
adjusted to the experimental energy
separation between the $s$ and $p$ $\Lambda$--levels in the hypernucleus.
The $\Lambda$ total energy in Eqs.~(\ref{decw}) and (\ref{decwpar3})
is given by $k_{0}=m_\Lambda+\v{k}^2/(2 m_\Lambda)+V_{\Lambda}$,
$V_\Lambda$ ($=-10.8$ MeV for $^{12}_\Lambda$C)
being a binding energy term.

One should keep in mind that, by definition, the decay widths cannot
contain any (final state) interaction among the weak decay nucleons \cite{BG10}. Nevertheless, in our approach the strong nucleon--nucleon interaction, $V^{N N}$, plays a role in the definition of the hypernuclear ground state, which, using perturbation theory up to second order in $V^{N N}$, reads:
\begin{eqnarray}
\label{gstate}
\ket{0}_{k_{F}} & = & \mathcal{N}(k_{F}) \, \Biggl(\ket{\;}
- \sum_{2p2h} \,
\frac{\bra{2p2h} V^{N N} \ket{\;}}
{E_{2p2h}-E_{HF}} \,
\ket{2p2h}  \\
&&  + \sum_{3p3h} \sum_{2p2h}\,
\frac{\bra{3p3h} V^{N N} \ket{2p2h} \; \bra{2p2h} V^{N N} \ket{\;}}
{(E_{3p3h}-E_{HF})(E_{2p2h}-E_{HF})} \,
\ket{3p3h} \Biggr)
\otimes \ket{\Lambda}~, \nonumber
\end{eqnarray}
where $\ket{\;}$ ($\ket{\Lambda}$) is the normalized and uncorrelated hypernuclear core ground state (hyperon initial state), i.e., the Hartree--Fock vacuum with energy $E_{HF}$, which we take equal to zero.
For the nucleon--nucleon interaction $V^{NN}$ we adopt a Bonn potential (with
the exchange of $\pi$, $\rho$, $\sigma$ and $\omega$ mesons) \cite{Ma87}. Note that
being $V^{N N}$ a two--body operator, it can connect $\ket{\;}$ only to a $2p2h$ configuration, as shown in the second and third terms in the rhs of the above expression. As our aim is to evaluate the decay rates up to $\Gamma_{3}$, we have restricted the third term in Eq.~(\ref{gstate}) to $3p3h$ final states. It is self--evident that $E_{2p2h}$ and $E_{3p3h}$ are the energies of the
$2p2h$ and $3p3h$ configurations, respectively. Moreover, $\mathcal{N}(k_{F})$ is the normalization function, which is easily evaluated by imposing $_{k_{F}}\langle 0 \ket{0}_{k_{F}}=1$.

The widths $\Gamma_1$, $\Gamma_2$ and $\Gamma_3$ are obtained
by inserting Eq.~(\ref{gstate}) into Eq.~(\ref{decw})
and then by performing the local density approximation through Eq.~(\ref{decwpar3}).
In what follows, we focus on $\Gamma_{3}$ (for a detailed
discussion on $\Gamma_{1}$ and $\Gamma_{2}$ we refer to~\cite{BG09,BG10}), for which one obtains:
\begin{eqnarray}
\label{decw3}
\Gamma_{3}(\v{k},k_{F})  & = & \mathcal{N}^{\, 2}(k_{F})
\sum_{f=4p3h} \, \delta (E_{f}-E_{0}) \left| \sum_{2p2h} \, \sum_{3p3h} \,
\bra{f} V^{\Lambda N\to nN} \ket{3p3h; \Lambda} \right. \nonumber \\
&&\times \left. \frac{\bra{3p3h} V^{N N} \ket{2p2h} \; \bra{2p2h} V^{N N} \ket{\;}}
{(E_{3p3h}-E_{HF})(E_{2p2h}-E_{HF})} \,\right|^{2}~. \nonumber
\end{eqnarray}

\begin{figure}[h]
\centerline{\includegraphics[scale=0.47, angle=-90]{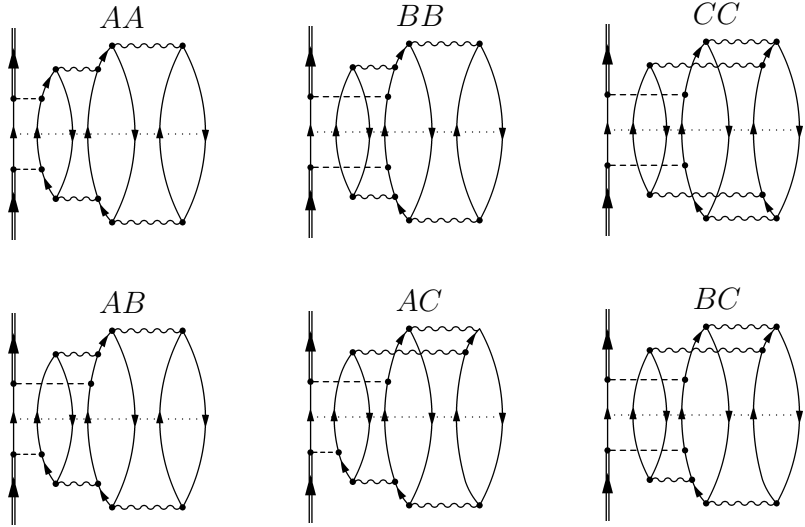}}
\caption{Goldstone diagrams contributing to the three--nucleon induced decay rate. The cuts indicating the $4p3h$ final states are also reported.}
\label{diagGoldstone}
\end{figure}

The Goldstone diagrams contributing to the three--nucleon induced decay mode are given in Fig.~\ref{diagGoldstone}.
Explicit expressions for these diagrams are obtained, as usual, using the standard rules for Goldstone diagrams. We remind the reader that our microscopic approach is inspired by the many--body scheme proposed by Oset and Salcedo in \cite{Os85} for the calculation of the one--nucleon stimulated decay rates and then extended to two--nucleon stimulated processes in \cite{Ra94,Ga00}. Although we adopt here a more realistic weak transition potential, the parts concerning the exchange of the $\pi$ and $\rho$ mesons are the same of \cite{Ra94,Ga00,Os85}.

{\bf \em Results} --
In Table \ref{gamm0} we give, for $^{12}_\Lambda$C, the contributions of the Goldstone diagrams of Fig.~\ref{diagGoldstone} to the various isospin channels of the three--nucleon induced decay mode.
The effect of nuclear recoil is not included in these predictions.
The $nnp$-- and $npp$--induced contributions dominate and are of the same order of magnitude for most of the diagrams.
The interference terms, $AB$, $AC$ and $BC$, some of which are negative, provide small contributions. Neglecting them, the total three--nucleon induced rate would be $\sim$ 5\% smaller than the full result, $\Gamma_3=8.209 \times 10^{-2}$ in units of the free $\Lambda$ decay rate.
For the complete calculation we obtain (see last line in Table \ref{gamm0}),
$\Gamma_{nnn}$ : $\Gamma_{nnp}$ : $\Gamma_{npp}$ : $\Gamma_{ppp}$ =
1 : 31 : 40 : 2.4.
The dominance of the $nnp$-- and $ppn$--induced decay modes originates from the relative importance of the different isospin channels in a twofold way: it is due to the isospin factors
and the way in which the weak and the strong interactions weight these factors in Eqs.~(\ref{decw}) and (\ref{gstate}). It turns out that the exchange of vector mesons provide dominant $nnp$-- and $npp$--induced contributions by the charge--exchange terms. The scalar mesons have smaller isospin factors and, together with the weights given by the different interactions, softens the dominance behavior by the vector mesons.
\begin{table}[h]
\begin{center}
\caption{Predictions for the isospin contributions to the three--nucleon induced decay mode of $^{12}_\Lambda$C originating from the diagrams of Fig.\ref{diagGoldstone} (in units of $10^{-2}$ times the free $\Lambda$ decay rate).}
\label{gamm0}
\begin{tabular}{cccccc}   \hline\hline
Diagram & $\Gamma_{nnn}$ & $\Gamma_{nnp}$ & $\Gamma_{npp}$ & $\Gamma_{ppp}$ & $\Gamma_{3}$\\ \hline
$AA$       & $0.038$      & $1.367$       & $1.281$ & $0.097$       & $2.783$        \\
$BB$        & $0.037$      & $1.064$       & $1.448$ & $0.091$       & $2.640$        \\
$CC$        & $0.032$      & $0.825$       & $1.422$ & $0.081$       & $2.360$       \\
$AB$        & $0.008$      & $0.116$       & $0.135$ & $0.005$       & $0.264$       \\
$AC$        & $-0.002$     & $0.011$       & $0.070$ & $-0.004$      & $0.075$       \\
$BC$        & $-0.003$     & $0.051$       & $0.047$ & $-0.008$      & $0.087$       \\ \hline
sum         & $0.110$       & $3.434$       & $4.403$ & $0.262$       & $8.209$      \\
\hline\hline
\end{tabular}
\end{center}
\end{table}
In Table \ref{g3mesons} the full result for $\Gamma_3$ is compared with the
partial predictions obtained by limiting the weak transition potential to 1) one--pion--exchange and 2) $(\pi+K)$--exchange. As expected, the $\pi$--meson provides the largest contribution to the decay width, while the effect of the $K$--meson is to reduce the rate.
\begin{table}[h]
\begin{center}
\caption{Effect of the weak transition potential on the total three--nucleon stimulated decay width.
The predictions for $\pi$-- and $(\pi+K)$--exchange are compared with the full result including the exchange of all mesons.
Units are the same as in Table~\ref{gamm0}.}
\label{g3mesons}
\begin{tabular}{lc}   \hline\hline
   ~~~~~Weak Potential~~~~~ &  ~~~~~$\Gamma_{3}$~~~~~   \\ \hline
$\pi$  &   $6.841$\\
$\pi+K$  &   $5.755$\\
$\pi+\eta+K+\rho+\omega+K^{*}$  &   $8.209$\\
\hline\hline
\end{tabular}
\end{center}
\end{table}

It is evident from Fig.~\ref{diagGoldstone} that we have restricted ourselves to the evaluation of direct diagrams in which the weak and strong interactions are attached to particles lines. We have neglected diagrams in which at least one of these interactions is connected to a hole line as well as Pauli exchange diagrams. The contributions of the ``hole diagrams'' are known to be small due to phase space considerations; their net effect would be an increase in the
value for the three--nucleon induced decay width. On the contrary, exchange terms would reduce the value of $\Gamma_3$. A rough estimate of the neglected diagrams can be made by using our prior knowledge on the two--nucleon induced mechanism. In \cite{BG09} we evaluated
the full set of diagrams for $\Gamma_{2}$ by using the same microscopic approach and weak and strong potentials of the present calculation. The reduction due to the exchange terms turns out to be stronger than the increase due to the neglected hole diagrams, resulting in a net decrease of $\Gamma_2$ by $\sim 20\%$. We expect a reduction of the same order of magnitude for $\Gamma_{3}$ once hole and exchange diagrams are taken into account. The evaluation of these effects is quite involved and goes beyond the scope of the present contribution.

Nuclear recoil is particularly easy to implement within our nuclear matter formalism, as we work in the momentum space. In particular, the sum over the final states in Eq.~(\ref{decw}) contains a summation over the momenta of the outgoing nucleons
$\vec{p}_{j}$, where $j=1,\dots,i+1$ for the rate $\Gamma_{i}$ ($i=1,2,3$). Momentum conservation is expressed as: $\vec{p}_{1} +\dots+ \vec{p}_{i+1} =-\vec{P}_{T}$, $\vec{P}_{T}$ being the momentum of the residual nucleus. The final nucleons are free particles; if we also assume that the residual nucleus is in its ground state, all the final state particles are characterized by their kinetic energy. The effect of nuclear recoil
on the decay rates thus results from subtracting the kinetic energy of the residual nucleus from the $\Lambda$ total energy $k_{0}$, thus replacing $k_{0}$ with $k_{0} - \v{P}_{T}^{2}/2 M_{\rm res}$, $M_{\rm res}$ being the mass of the recoiling nucleus.
In Table \ref{recoil} we show the effect of the recoil on the decay widths $\Gamma_1$, $\Gamma_2$ and $\Gamma_3$. It tends to decrease, on average, the momenta of the final nucleons, which are then more Pauli blocked.
We indeed found a reduction effect due to the recoil, which is particularly relevant for $\Gamma_3$ ($\sim$ 15\%), less important for $\Gamma_2$ ($\sim$ 4\%) and negligible for $\Gamma_1$; for the total nonmesonic rate $\Gamma_{\rm NM}$ the decrease is $\sim$ 3\%. As expected, the reduction of the final nucleon momenta, i.e., of the decay rate, increases with the number of nucleons produced in the decay.
Note also that the rate $\Gamma_{1}$ is strongly dominated by the back--to--back kinematic for the final nucleon pair, i.e., by small values of $|\v{P}_{T}|$. At variance, nucleons are emitted with almost no preferential relative directions in two-- and three--nucleon induced decays, resulting in values of $|\v{P}_{T}|$ which can be significant and producing non--vanishing reductions of $\Gamma_{2}$ and (especially) $\Gamma_{3}$.
\begin{table}[h]
\begin{center}
\caption{Effect of the recoil of the residual nucleus on the one--, two-- and three--nucleon stimulated decay widths of $^{12}_\Lambda$C.
Results are given in units of the free $\Lambda$ decay rate.}
\label{recoil}
\begin{tabular}{lcccc}   \hline\hline
  & ~~~~~$\Gamma_{1}$~~~~~ & ~~~~~$\Gamma_{2}$~~~~~ & ~~~~~$\Gamma_{3}$~~~~~ & ~~~~~$\Gamma_{\rm NM}$~~~~~ \\ \hline
without        & $0.601$      & $0.301$       & $0.082$  & $0.984$\\
with            & $0.600$      & $0.288$       & $0.070$  & $0.958$\\
\hline\hline
\end{tabular}
\end{center}
\end{table}
Another comment is in order on the final results for $\Gamma_{1}$, $\Gamma_{2}$ and $\Gamma_{3}$ of Table \ref{recoil}.
The average momentum for the nucleons emitted in three--nucleon induced decays is $\sim$ 290 MeV$/c$, a value which is only slightly larger than the Fermi momentum of nuclear matter, $k_F\sim$ 270 MeV$/c$. The Pauli blocking on these nucleons is thus severe and contributes to the small value of the ratio $\Gamma_3/\Gamma_1$, which turns out to be $\sim$ 0.12. Moreover, our predictions implies that
$\Gamma_2/\Gamma_1(\sim0.48)>\Gamma_3/\Gamma_2(\sim0.24)$, which indicates that decays induced by four or more nucleons are expected to be negligible.

In Table \ref{res-gamas} we present the comparison of our final results for the full set of $^{12}_\Lambda$C decay widths with the latest KEK \cite{KEK09} and FINUDA \cite{FINUDA3,FINUDA4} data.
\begin{table}[h]
\begin{center}
\caption{Predictions and recent data from KEK--E508 \cite{KEK09} and FINUDA \cite{FINUDA3,FINUDA4} for the nonmesonic weak decay widths of $^{12}_\Lambda$C (in units of the free $\Lambda$ decay rate). See text for details.}
\label{res-gamas}
\begin{tabular}{c c c c c}
\hline\hline
 & \mc{1}{c} {Our} & {KEK--E508} \cite{KEK09} &  {FINUDA \cite{FINUDA3}}  &
{KEK--FINUDA \cite{FINUDA4}}
\\ \hline
$\Gamma_n$ & $0.145$ & $0.23\pm0.08$ & $$ & $0.28\pm0.12$\\
$\Gamma_p$ & $0.455$ & $0.45\pm0.10$ & $0.65\pm0.19$ & $0.493\pm0.088$  \\
$\Gamma_1$ & $0.600$ & $0.68\pm0.13$ & $$ & $0.78\pm 0.09$\\
$\Gamma_2$ & $0.288$ & $0.27\pm0.13$ & $$ & $0.178\pm0.076$\\
$\Gamma_3$ & $0.070$ & $-$ & $-$ & $-$ \\
$\Gamma_{\rm NM}$ & $0.958$ & $0.953\pm0.044$ & $$ & $0.96\pm0.04$  \\
$\Gamma_n/\Gamma_p$ & $0.319$ & $0.51\pm0.14$ & $$ & $0.58\pm0.27$\\
                                     &            & $0.29\pm0.14$ \cite{GPR03} & $$ & \\
                                     &            & $0.34\pm0.15$ \cite{BGPR10} & $$ & \\ \hline
$\Gamma_1/\Gamma_{\rm NM}$ & $0.626$ & $0.71\pm 0.14$ & $$ & $0.81\pm0.10$ \\
$\Gamma_2/\Gamma_{\rm NM}$ & $0.301$ & $0.29\pm0.13$ &
$0.25\pm0.12\pm0.02$ & $0.19\pm0.08$\\
 & $$ & $$ & $0.20\pm0.08^{+0.04}_{-0.03}$ &  $$\\
$\Gamma_3/\Gamma_{\rm NM}$ & $0.073$ & $-$ & $-$  & $-$\\
\hline\hline
\end{tabular}
\end{center}
\end{table}
A few details on the experiments are needed for a better understanding of the comparison. The first (second) FINUDA determination of $\Gamma_2/\Gamma_{\rm NM}$ and the data for $\Gamma_p$ refer to proton (neutron--proton correlation) spectra analyses. The determinations listed as KEK--FINUDA have been reconstructed in the FINUDA paper \cite{FINUDA4} by starting from various existing KEK data on $\Gamma_p$ and the total ($\Gamma_{\rm T}$) and mesonic rates ($\Gamma_{\rm M}$), together with FINUDA values for $\Gamma_p$ and $\Gamma_2$. The KEK--FINUDA rate $\Gamma_n$ ($\Gamma_1$ and $\Gamma_{\rm NM}$) is obtained as the difference $\Gamma_n=\Gamma_{\rm T}-\Gamma_{\rm M}-\Gamma_p-\Gamma_2$ ($\Gamma_1=\Gamma_{\rm T}-\Gamma_{\rm M}-\Gamma_2$ and $\Gamma_{\rm NM}=\Gamma_{\rm T}-\Gamma_{\rm M}$). Moreover, the KEK--FINUDA result for $\Gamma_p$ is nothing but the weighted average between the KEK and FINUDA determinations.
The KEK determination $\Gamma_n/\Gamma_p=0.51\pm0.14$ was considered to be free from ambiguities due to nucleon final state interactions and two--nucleon induced decays in \cite{KEK09}. Actually, these processes are responsible for a non--negligible reduction of the experimental value of $\Gamma_n/\Gamma_p$ \cite{Rev-hypernuclei,Rev-decay}: the determination $\Gamma_n/\Gamma_p=0.29\pm0.14$ ($0.34\pm0.15$) shown in Table \ref{res-gamas} is a theoretical fit of KEK coincidence $nn$ and $np$ spectra obtained within the present microscopic approach in \cite{GPR03} (a finite nucleus calculation in \cite{BGPR10}) by including the mentioned many--body effects. These effects lead to the inequality $\Gamma_n/\Gamma_p< N_{nn}/N_{np}=
0.51\pm0.14$.

An overall agreement of our predictions with the experiments is evident from Table \ref{res-gamas}, especially with KEK data (it is of the same quality as the comparison between the central values of KEK and FINUDA data). Note how the agreement is especially good for the most easily measurable rate, $\Gamma_{\rm NM}$, while for $\Gamma_n$, the rate affected by the largest error bars together with $\Gamma_2$, the agreement is worse. The prediction for the $\Gamma_n/\Gamma_p$ ratio compares well with the fits of \cite{GPR03,BGPR10} to KEK spectra. The only discrepancies which exceeds the level of $1\sigma$ are with the KEK--FINUDA rates $\Gamma_n$ (and then $\Gamma_1$) and $\Gamma_2$ and the KEK $\Gamma_n/\Gamma_p$ ratio obtained without including the two--nucleon induced mode and final state interaction (see the discussion of the previous paragraph on the experimental determinations of $\Gamma_n$ and $\Gamma_n/\Gamma_p$).

The rate $\Gamma_3$ amounts to $\sim$ 7\% of the total nonmesonic rate $\Gamma_{\rm NM}$ and is about a half of the neutron--induced rate $\Gamma_n$. Therefore, the three--nucleon induced contribution cannot be neglected if one aims at a detailed understanding of the full set of decay rates. We finally note that the KEK and KEK--FINUDA determinations of $\Gamma_1$ tend to overestimate the value predicted here; given the good agreement for $\Gamma_{\rm NM}$, one cannot exclude the eventuality that the KEK and FINUDA experiments counted a fraction of nucleons originating from three--nucleon induced decays (and from two--nucleon induced decays too, in the case of FINUDA) as wrongly produced by one--nucleon induced processes.

{\bf \em Conclusion} --
The first study of the three--nucleon stimulated nonmesonic weak decay of hypernuclei is presented within a microscopic approach. The full set of decay rates are obtained for $^{12}_\Lambda$C also including the effect of nuclear recoil. The three--nucleon induced rate $\Gamma_3$ turns out to be non--negligible
---amounting to  $\sim$ 7\% of $\Gamma_{\rm NM}$ and being $\Gamma_3/\Gamma_n \sim 1/2$---
and is dominated by the $nnp$-- and $npp$--induced channels. Nuclear recoil basically only affects $\Gamma_3$. The predictions show a good overall agreement with present KEK and FINUDA data.

In spite of the last advances in the field of nonmesonic decay, discrepancies between theory and experiment persist and mainly concern emission spectra involving protons. The results obtained in the present Letter for the three--nucleon stimulated decays can hardly improve this disagreement, as these decays are expected to provide almost the same proportion of weak decay, i.e., primary, neutrons and protons ($\Gamma_{nnp}\sim \Gamma_{npp}\sim \Gamma_3/2$).  However, note that additional, secondary nucleons coming from final state interactions (not included here in the calculation of the decay rates) must be taken into account in spectra analysis.
It is also worth mentioning that the quality of present data does not allow us to establish the degree of violation of the $\Delta I=1/2$ isospin rule in the one--nucleon induced nonmesonic weak decay \cite{Rev-decay,AG00}. Most of the adopted meson--exchange models, including the present one, only contain pure $\Delta I=1/2$ transitions.
Further work is thus necessary to achieve the primary purpose of hypernuclear weak decay studies, which is to access the properties of the elementary strangeness--changing hyperon interactions.
The efforts of the recent years clearly indicates that nuclear many--body effects cannot be disregarded nor easily disentangled from the basic weak decay ingredients; mechanisms such as nucleon final state interactions and multinucleon induced decays are relevant, they partially superimpose with one another and mask the elementary weak interactions.

{\bf \em Outlook} --
Systematic and more precise measurements of nucleon spectra, decay widths and asymmetries ---possibly over some mass number range, as in the spirit of FINUDA and under a close collaboration with theoreticians--- will be essential to move forward.
An approved proposal at J--PARC consists in the E18 experiment \cite{JparcE18}. It is designed to have much better statistics than KEK--E508 and concerns measurements of the rates $\Gamma_n$, $\Gamma_p$ and $\Gamma_2$ for $^{11}_\Lambda$B and $^{12}_\Lambda$C. Triple nucleon coincidence measurements could lead to (the first) direct measurements of $\Gamma_2$ with a 10\% statistical error. The kinematics of precise triple coincidence measurements could provide some evidence of the three--nucleon induced decay mode. Perhaps, an experiment with slightly improved setup and performances compared to those of E18 could be suitable for extending the direct observation to three--nucleon stimulated decays.
A second J--PARC approved experiment, E22 \cite{JparcE22}, consists in a high statistics study of the $\Delta I=1/2$ rule for $^4_\Lambda$H and $^4_\Lambda$He.
An indication for the possibility of new experiments, which could be performed at J--PARC by using the successful techniques developed by FINUDA, has been put forward in the last paper quoted in \cite{Rev-decay}. It consists in measurements of the full set of decay rates (including the total and the mesonic ones) for $^5_\Lambda$He, $^7_\Lambda$Li, $^9_\Lambda$Be, $^{11}_\Lambda$B, $^{12}_\Lambda$C, $^{15}_\Lambda$N and $^{16}_\Lambda$O with a statistical precision of $\sim$ 5\%. We also remind the proposal of \cite{FINUDA2016} for new measurements of the lifetimes and the proton--induced rates of $^3_\Lambda$H, $^4_\Lambda$H, $^{12}_\Lambda$B and other neutron--rich $p$--shell hypernuclei. We take the opportunity to strongly encourage the experimental colleagues to consider the feasibility of a measurement of the three--nucleon induced weak decay mode. From the theoretical side, our approach is particularly suitable for including new ingredients scarcely or never investigated before such as $\Delta I=3/2$ terms and the $\Lambda N$--$\Sigma N$ coupling.

{\bf \em Acknowledgments} --
This work was partially supported by the CONICET, Argentina, under contract PIP 00273.


\end{document}